\documentclass[12pt]{iopart}
\usepackage{bm} 
\expandafter\let\csname equation*\endcsname\relax
\expandafter\let\csname endequation*\endcsname\relax
\usepackage{iopams}
\usepackage{amsmath}
\usepackage{amssymb}
\usepackage{amsfonts}
\usepackage{mathrsfs}
\usepackage{comment}
\usepackage{graphicx}
\usepackage{color}
\usepackage{dcolumn} 
\usepackage{soul} 
\usepackage{cancel}
\usepackage[T1]{fontenc}
\usepackage[utf8]{inputenc}

\usepackage{comment} 

\usepackage{epstopdf} 

\begin{document}
\title{Intensity pseudo-localized phase in the glassy random laser}

\author{Jacopo Niedda$^{1,2}$, Luca Leuzzi$^{2,1,\dagger}$,  Giacomo Gradenigo$^{3,4}$}
\address{$^1$ Dipartimento di Fisica, Universit\`a di Roma
  ``Sapienza'', Piazzale A. Moro 2, I-00185, Roma, Italy}

\address{$^2$ NANOTEC CNR, Soft and
  Living Matter Lab, Roma, Piazzale A. Moro 2, I-00185, Roma, Italy}

\address{$^3$ Gran Sasso Science Institute, Viale F. Crispi 7, 67100 L'Aquila, Italy}
\address{$^4$ INFN-Laboratori Nazionali del Gran Sasso, Via G. Acitelli 22, 67100 Assergi (AQ), Italy}

\ead{$^\dagger$ luca.leuzzi@cnr.it}

\begin{abstract}

 Evidence of an emergent pseudo-localized
phase characterizing the low-temperature replica symmetry breaking phase
of the complex disordered models for glassy light is provided in the mode-locked random laser model. A pseudo-localized
phase  corresponds to a state in which the intensity of light modes is neither
equipartited among all modes nor strictly condensed on few of them.
Such a \emph{hybrid} phase, recently characterized as a finite size effect in other
models, such as the Discrete Non-Linear Schr\"odinger equation, in the low temperature phase of the glassy random laser appears to be robust in the limit of large size.
\end{abstract}

\maketitle

\section{Introduction}

The present paper is devoted to the study of a particular phenomenon
taking place in quenched disordered models of random lasers~\cite{Angelani06a,Antenucci15a}: intensity localization of
light, else termed \emph{power condensation} \cite{Antenucci15c,Antenucci15d}, and its relationship to the high pumping 
replica symmetry breaking (RSB) random lasing phase.

It was since the seminal work of Anderson on semiconductors
that localization {\color{black}{(space localization in this case)}} has been related to ergodicity breaking~\cite{Anderson58}.
In his most celebrated work Anderson showed that a large enough amount
of disorder, in the form of impurities of the alloy, was able to
induce the localization in space of the electron wavefunction and the consequent
absence of transport,  therefore triggering a semiconductor/insulator
transition.
The insulating-localized phase was clearly
characterized by a dynamical breaking of~\emph{ergodicity}. 

Since the'50s the idea of localization, related to ergodicity breaking, has been conceived also in
classical systems: we think of the celebrated Fermi-Pasta-Ulam-Tsingou
(FPUT) numerical study on the relaxation from an
atypical initial condition in a slightly anharmonic chain. Despite
Fermi's expectation that the anharmonicity was sufficient to trigger
the fast relaxation of the system to equipartion between the Fourier
modes of the chain, an initial condition with only the lowest harmonic
excited showed a recurrent dynamics for the whole duration of the
numerical experiment, with no signature of relaxation to
equipartition~\cite{Fermi55}. Localization was, thus, coming along with dynamical ergodicity breaking also in
the case of energy localization in the Fourier power
spectrum.

The paradigm of ergodicity breaking as a dynamical phenomenon
manifesting itself as the lack of transport or the absence of
relaxation from atypical initial conditions has been the dominant one
until the '80s, when a thermodynamic paradigm and a statistical
ensembles formalism was established for ergodicity breaking
transitions in complex disordered systems: {\color{black}{the theory of replica symmetry breaking proposed by
Parisi~\cite{Parisi79,Mezard87}.~Quite interestingly, while localization
phenomena in quantum many-body systems have been widely investigated
during the last decades~\cite{Altshuler97,Basko06,Nandkishore15,Ros15,Vidmar16,Abanin17,Alet18,Mossi18},
they have seldom been probed in disordered systems, apart from some attempts, e.g., Refs.~\cite{Gradenigo21a,Gradenigo21b}. }} 
This lack of analysis in disordered glassy systems is due to the nature of the variables which are customary for magnetic disordered systems: Ising, XY or 
Heisenberg spins are all locally bounded, $|\vec s_i|=1 , ~\forall ~i=1,\ldots, N$, with $N$ being the size of the system.
In those models where variables were used with continuous locally unbounded magnitude as a proxy for magnetic spins in spin-glasses or density fluctuations in structural glasses \cite{Kirkpatrick87b,Kirkpatrick89,Crisanti92,Crisanti11,Sun12,Crisanti13},
the interaction network was fully connected, thus hindering any sort of magnitude localization.~Indeed,
this kind of mean-field representation on a complete graph, together with a local potential (soft spins) or a global constraint (spherical spins) guarantees magnitude equipartition.

A careful investigation of how  magnitude localization
coexists with replica symmetry breaking in disordered systems is
therefore a gap that needs to be filled. Some signatures of strong breaking of intensity equipartition  have been already
highlighted in Ref.~\cite{Gradenigo20} as features of the emission spectra of glassy random lasers at high pumping \cite{Folli12,Folli13,Ghofraniha15,Gomes16,Basak16,Pincheira16,Gomes21}. 
By means of numerical simulations on a slightly different model for random lasers with respect to the one of Ref.~\cite{Gradenigo20}, we will now deepen the study of such phenomenology  and we will  connect it with the most
recent literature on intensity localization in systems with locally unbounded
variables and global constraints \cite{Gradenigo21a,Gradenigo21b,Franzosi11,Rasmussen00,Iubini13}.

The leading model for the non-linear interaction of light modes in a random optically active medium (under external pumping) is the so-called ``mode-locked $p$-spin", a model with complex spherical spins and mixed $p-$body potential with both the $p=2$ and $p=4$ interactions~\cite{,Gradenigo20}. 
The mode-locked $p$-spin differs from the original fully connected model because not all the possible $p$-uples of light modes (each mode being identified by his angular frequency $\omega_k$ and its complex amplitude $a_k=a(\omega_k)$) participate to the $p$-body interaction, only those allowed by the symmetries of light-matter interaction in random active media. It is only in the \emph{narrow-bandwidth} approximation that all  $p$-uplets are allowed to interact and contribute to the Hamiltonian. In this case the model has been analytically solved in the framework of Replica Symmetry Breaking theory \cite{Antenucci15a,Antenucci15b}. 
In the narrow-bandwidth approximation  each light mode interacts with $\mathcal{O}(N^3)$ modes ($p=4$) and, despite the fact that the continuous variables of the model, i.e.~light modes,  
are locally unbounded, the  complete, fully connected,  character of the interaction network  forbids any sort of inhomogeneous distribution of intensity. 
Full connectivity enforces intensity equipartition among light modes. On the other hand, in a diluted  network the overall magnitude might be subdivided among variables in a non-homogeneous way and, if the dilution is strong enough, magnitude localization might occur.

 We stress that {\color{black}{power condensation (else termed intensity, or magnitude, localization)}}   is not the generalization to light waves of the spatial wavefunction localization
occurring in  Anderson theory, that is known to be inhibited in 3D random lasers because of the vectorial nature of light waves \cite{Skipetrov14,Sperling16}.
Rather, it is a condensation of the overall magnitude on a few Fourier modes, in  presence of a global constraint on intensity.
When the value of this globally conserved
quantity exceeds a given  threshold the system
undergoes a  transition where a macroscopic fraction of
the conserved quantity concentrates, {even in the thermodynamic limit, on a finite set of degrees of freedom (Fourier modes in the present case)}.

As an instance, this behavior has been found and very precisely described
in the framework of large deviation calculations and ensemble
inequivalence in the case of mass-transport models~\cite{MEZ05,EMZ06}
or for bosonic condensates in optical lattices described by the
Discrete Non-Linear Schr\"odinger Equation (DNLSE)~\cite{TS01}.
Concerning the DNLSE, it is known since almost two decades that
his high energy phase is characterized by the appearance of localized breather-like solutions, which typically arise in certain regimes in models of non-linear
waves~\cite{K09,LFO06,Franzosi11,Iubini21,Gotti21,Politi22,Iubini14}. 

A complete understanding of the relevance of these solutions from the point of view of equilibrium statistical mechanics has been achieved only recently: in Refs.~\cite{Gradenigo21a,Gradenigo21b} it has been shown
that they are stable equilibrium solutions in the microcanonical
ensemble {with the peculiarity of having a} negative temperature~\cite{Rasmussen00,Iubini12,Iubini13,Iubini17,Gotti21}. The presence of a negative temperature, which in the context of the DNLSE comes along with the lack of statistical ensemble equivalence, is a signature of a ``thermodynamic anomaly": the larger the amount of energy fed to the system,
the larger the amount of condensate mass localization, hence the smaller the entropy. Such a phenomenon is usually not possible in the canonical ensemble but is perfectly allowed in the microcanonical one.

The very interesting feature of the magnitude localization transition taking
place for bosonic condensates on a lattice is the existence of a pseudo-localized
phase, which is a sort of precursor of the localized one. This is interesting also in connection to the high pumping behavior of mode-locked random lasers.
In the case of the bosonic condensates the pseudo-localized phase is  a \emph{hybrid} phase: the participation ratio (the localization order
parameter) vanishes for a large number $N$ of degrees of freedom, but it displays  a
negative temperature, that is a signature of incipient localization. A negative temperature signals a non-standard
thermodynamic behaviour, where the entropy decreases by increasing the
energy, and a breaking of equivalence between the microcanonical and the canonical ensembles in statistical mechanics. 
As we are going to discuss, in the high pumping replica symmetry breaking phase of the mode-locked complex spherical $p$-spin model another hybrid pseudo-localized phase occurs: for large systems
the participation ratio goes to zero but the emission intensity spectra are very inhomogeneous.

 An important difference between the two cases is that in the DNLSE condensates the pseudo-localization is related to the
inequivalence between canonical and microcanonical ensembles, while in
the present case it is not. 
This comes about because in the laser model the hard constraint is not on the energy (as in the microcanonical ensemble) but, rather, on the overall intensity, that does not coincide with the energy. 
Indeed, in the present work the emergence of a  pseudo-localized phase occurs in the canonical ensemble and it can be studied with ordinary (though highly optimized) Monte Carlo algorithms.


We will see that a sensitive order parameter for the pseudo-localization in the mode-locked random laser 
is related to the spectral entropy signaling the breaking of intensity equipartition occurring at some critical pumping power. 
This critical point is consistent with the RSB transition point, in the numerical uncertainty of the finite size scaling for the simulated systems. We will show that a fundamental issue in the occurrence of localization or  pseudo-localization of intensity in the low temperature/high pumping phase is the connectivity of the nonlinear mode interaction.

\section{Model and Objectives} \label{model}

The leading model for the mode-locking random laser is \cite{Antenucci15a,Antenucci16,Antenucci16b,Gradenigo20}
 \begin{eqnarray}
  \label{Hamilt1}
 \mathcal{H}[\bm{a}] = 
 - \sum_{\bm{k} | {\rm FMC}(\bm{k})} J_{k_1 k_2 k_3 k_4} \overline{a}_{k_1}a_{k_2}\overline{a}_{k_3}a_{k_4} + \mbox{c.c.} ,
 \end{eqnarray} 
where the ``spins'' $a_k$ are the complex amplitudes of the light modes of angular frequency $\omega_k$, satisftying the global ``spherical'' constraint
\begin{equation} 
  \sum_k |a_k|^2 = \mathcal E= \epsilon N,
  \label{spherConstr}
\end{equation}
where $\epsilon = \mathcal{E}/N$ is the optical power per mode
available in the system and
the quenched disordered couplings $J_{k_1 k_2 k_3 k_4}$ are distributed according to 
\begin{equation} 
\mathcal{P}(J_{k_1 k_2 k_3 k_4}) = \frac{1}{\sqrt{2 \pi \sigma_4^2}} \exp\left\{-\frac{J^2_{k_1 k_2 k_3 k_4}}{2 \sigma_4^2}\right\},  \label{Gaussian}
\end{equation}
where 
\begin{equation}
   \sigma_4^2 = \frac{4! J_4^2}{2 N^2}. 
\end{equation}

The modes involved in each interaction term must satisfy the frequency matching condition (FMC) 
\begin{equation} \label{FMC}
{\rm FMC}(\bm{k}): | \omega_{k_1} - \omega_{k_2} + \omega_{k_3} - \omega_{k_{4}} | \lesssim \gamma,
\end{equation}
where $\gamma$ is the  single mode linewidth. In the simulated system 
the mode frequencies are chosen according to the simplifying scheme of being ``comb-like" distributed, 
\begin{equation}
    \omega_k = \omega_0 + \delta\omega~k,
\end{equation}
so that the FMC conditions simply read as
\begin{equation}
    |k_1-k_2+k_3-k_4| = 0 \quad , k_i=1,\ldots, N.
\end{equation}

{\color{black}{At difference with the model studied in Ref. \cite{Gradenigo20} we impose periodic boundary conditions (PBC) on the mode-frequencies, i. e., given any two mode indices $k_{a}$ and $k_b$ their distance wll be
\begin{eqnarray}
\nonumber
|k_a-k_b| = \left\{ \begin{array}{c c} 
|k_a-k_b| & \mbox{if } |k_a-k_b|\leq \left[\frac{N}{2}\right]
\\ 
&
\\
N-|k_a-k_b|
& \mbox{if } |k_a-k_b|\geq \left[\frac{N}{2}\right]
\end{array}\right. ,
\\
\label{E-FPBC}
\end{eqnarray}
where $[N/2]$ is the integer part of $N/2$.
These conditions on the space of frequencies have been introduced in Ref.~\cite{Niedda22} to reduce the strong finite-size effects on the numerical results of the ML 4-phasor model dynamics \cite{Gradenigo20}.
In practice, band-edge modes participate in the same number of interacting quadruplets  as the modes in the center of the spectrum. 
Therefore, the simulated model at a certain size $N$ can be regarded as the bulk of the  model with free boundaries on the mode frequencies and a  size larger than $N$. 
}}

The disorder-dependent partition function is 
\begin{equation}
    \label{eq:ZRL}
    Z_N(\beta,\mathcal E) = \int \prod_{k=1}^N da_k d\bar a_k \ e^{- \beta \mathcal H[\bm a]}\delta\left( \mathcal E- \sum_{k=1}^N |a_k|^2 \right),
\end{equation}
where $ \beta$ is proportional to the inverse heat bath temperature.~The Hamiltonian \eqref{Hamilt1} is compatible with the ``strong-cavity" limit \cite{Conti11}, where the $2$-body term of the general model mentioned in the introduction can be neglected because of   the spherical constraint. This is justified because we are mainly interested in the study of systems near and above the lasing transition, where the nonlinear $4$-body term is dominant. We refer to the model \eqref{Hamilt1} as mode-locked (ML) 4-phasor model.
 
 \subsection{Connectivity}
\label{ss:connectivity}
The variance $\sigma_4^2$ of the distribution (\ref{Gaussian}) is of order $N^{-2}$ to guarantee the extensivity of the energy of the thermodynamic states, given by the Hamiltonian (\ref{Hamilt1}). In fully connected $4$-spin interacting models the variance  is of $\mathcal O(N^{-3})$, but the condition (\ref{FMC}) turns out to dilute the complete graph of interaction of an order $N$ \cite{Marruzzo18}.
As already mentioned, it is precisely the dilution  introduced by the
frequency matching condition that allows for the occurrence of such heterogeneity in the mode intensities to manifest themselves in the form
of a pseudo-localized phase arising at the RSB transition.  

\begin{figure}[b!]
\center
\includegraphics[width=.7\columnwidth]{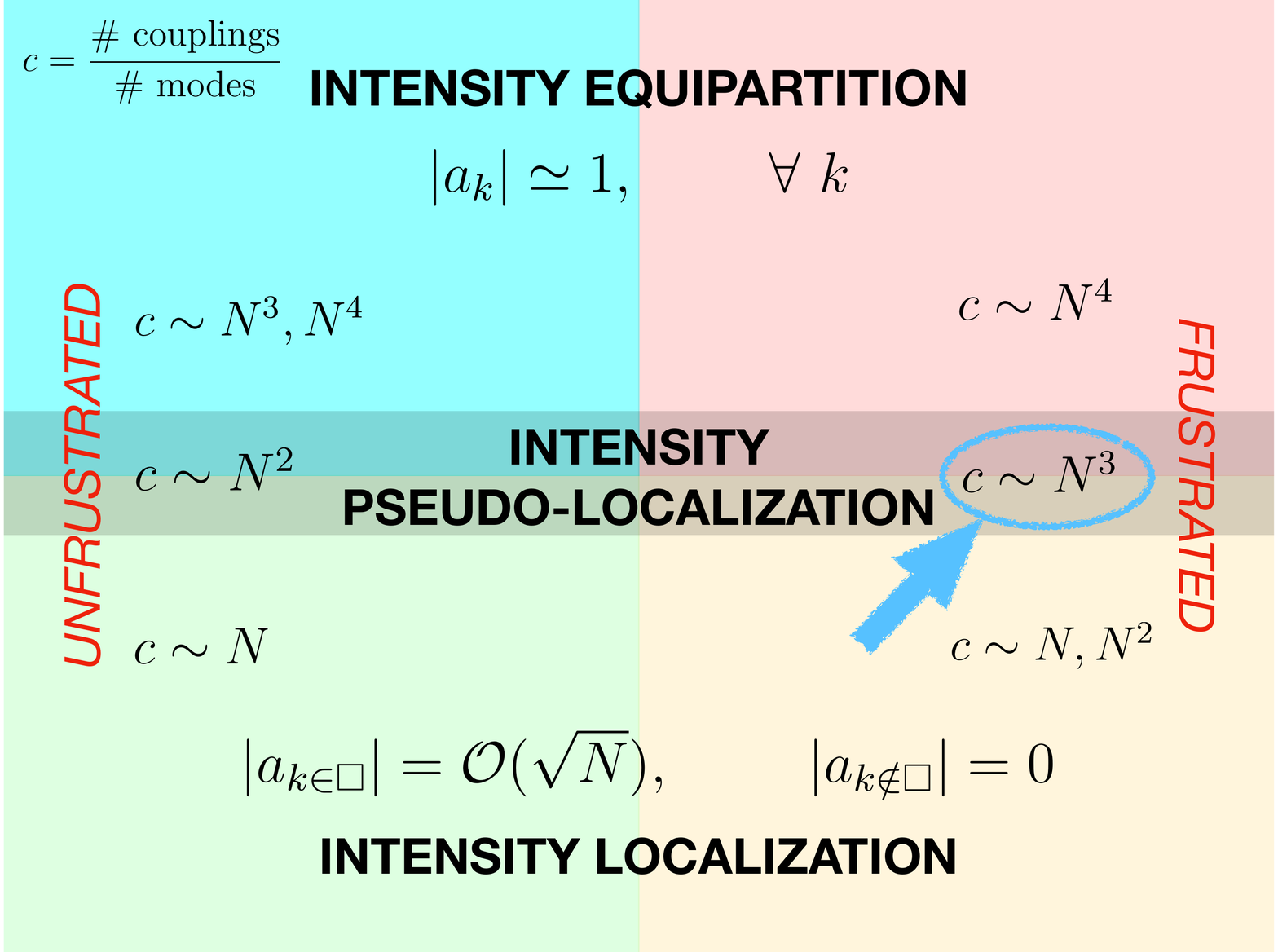}
\caption{Schematic representation of the possible regimes of low temperature intensity equipartition, intensity localization or pseudo-localization with respect to the coupling dilution in the $4$-spin spherical model, both for the  unfrustrated version (left) and the frustrated, strongly disordered, one (right). The case of the mode-locked random laser whose connectivity $c \sim N^3$ is pointed at by the arrow.}
\label{fig:dilution}
\end{figure}

In appendix \ref{app:loc_vs_equ} we go through a scaling argument on generic $p$-spin interacting systems (both with ordered or disordered interactions) with continuous variables and a global spherical  constraint, of which the ML $4$-phasor model with the constraint (\ref{spherConstr}) is a special case. 
In figure (\ref{fig:dilution}) we summarize the scaling argument predictions on a pictorial diagram with the known cases of equipartition, localization and pseudo-localization for the $4$-spin model.
There, on the left side, we include the ordered case, in which 
the couplings $J$ are not random at all, and the disordered but not frustrated case, in which the $J$'s are 
random but their mean square displacement is so small in comparison to the average that frustration does not take place. 
When frustration dominates, the low temperature/high pumping regime is glassy. This is the case depicted on the right hand side of the figure (\ref{fig:dilution}). The present model case has random couplings with zero mean, therefore guaranteeing frustration and glassiness, and an overall connectivity of order $N^3$. It is pointed out at by an arrow.
To summarize, in a $p$-spin interacting model, such as the one we are analyzing here, the external parameter driving the transition from an equipartite to a localized regime (passing through a ``critical'' pseudo-localized phase) is the order of dilution of the mean spin connectivity.

\subsection{Hard and soft constraints}
The
other fundamental ingredient involved in this phenomenology is the
spherical constraint (\ref{spherConstr}). 
In order to understand its role and importance
let us draw the parallelism with other models where localization takes
place. Let us consider for instance the DNLSE whose partition function reads
\begin{equation}
  \Omega(\mu,E) = \int \prod_{i=1}^N d\psi_id\bar \psi_i \ e^{-\mu \sum_{i=1}^N |\psi_i|^2}~\delta\left(E - \sum_{i=1}^N |\psi_i|^4\right),
  \label{eq:semican-bec}
\end{equation}  
where $\mu$ is the chemical potential. Clearly the quantity
\begin{equation}
  A = \sum_{i=1}^N |\psi_i|^2,
  \label{eq:spheric-DNLS}
\end{equation}
represents the mass of the condensate, so that the partition function in Eq.~\eqref{eq:semican-bec} corresponds to an ensemble where exact conservation of energy is enforced by means of a hard constraint
whereas mass is conserved only on average by means of a soft constraint.


In the case of the DNLSE, as in \textit{all}
cases where it takes place, the physical quantity that localizes is the
one controlled by the hard constraint, hence in this case it is the
energy. It is only thanks to the \textit{global} action of the
constraint on the total energy that configurations with a strongly
heterogeneous distribution of energy on lattice sites are allowed. The analytical calculations
of Refs.~\cite{Gradenigo21a,Gradenigo21b} show that as soon as energy is constrained
above a certain critical value, $E>E_c$, these localized
configurations dominate the partition function.~It can be shown
analytically, but it is also easy to make sense of it looking at
Eq.~\eqref{eq:semican-bec}, that one cannot have localization of a
quantity controlled ``on average''. It is not possible to have
strongly inhomogeneous fluctuations and/or localization of something
which is controlled \textit{homogeously} by means of a Lagrange
multiplier like the chemical potential in Eq.~\eqref{eq:semican-bec}.
This is precisely the same mechanism of Bose-Einsten condensation,
which is a form of localization in Fourier space: the
condensed phase can not be reached by controlling density with a chemical potential; density 
must be tuned directly, for instance decreasing the volume for a given
number of bosons~\cite{Huang87}.

The fact that the 
Bose-Einstein condensation cannot be implemented by tuning the chemical potential of a
reservoir in contact with the system is  analogous to the
fact that energy localization cannot be
achieved  by tuning the temperature of a
thermostat, i.e.~by studying the partition function where 
conservation of energy is imposed ``on average'', as $\exp(-\beta
\mathcal{H})$,  rather than exactly, as $\delta(E-\mathcal{H})$. In both the above
examples localization entails lack of statistical ensemble
equivalence: for the energy localization in the DNLSE it is the lack of
equivalence between fixed temperature and fixed energy ensembles, for
Bose-Einstein condensation it is the lack of equivalence between fixed
chemical potential and fixed density ensembles.

What about our random laser? In 
the partition function of
Eq.~\eqref{eq:ZRL} the
conservation of energy is realized on average, by imposing
homogeneously a temperature for all interacting quadruplets, while the
conservation of the total intensity is realized exactly, by means of the 
hard global constraint (\ref{spherConstr}). It is then possible to guess
that in the ML $4$-phasor model the intensity localization might occur, rather than energy. 
As for the condensation in DNLSE,
intensity localization is achieved by tuning the physical quantity $\mathcal E$ controlled by
the overall hard constraint  (\ref{spherConstr}). More precisely, when a certain threshold is overcome in the controlling parameter of the constraint, $\mathcal E>\mathcal E_c$,
configurations where a finite amount of the overall intensity is stored in a few, $\mathcal{O}(1)$, quadruplets might become thermodynamically dominant.

Actually, it is known from the theory \cite{Antenucci15b} that the external parameter driving the phase transition is the pumping rate 
\begin{equation}
    \mathcal P\equiv \frac{\epsilon}{\sqrt{T}},
    \label{eq:spherical-ok}
\end{equation} $T$ being the heat-bath temperature and $\epsilon=\mathcal{E}/N$ being the average intensity per mode defined in Eq. (\ref{spherConstr}). Therefore, $\mathcal P$ can be tuned either by varying $\epsilon$ at constant temperature (as experiments are usually carried out, apart from some exceptions \cite{Wiersma01,Zhai10}), or, equivalently, by changing $T$. In order to numerically simulate the model with Monte Carlo algorithms, one can sample equilibrium configurations at different values of $\beta=1/T$ from the canonical probability distribution associated to the partition function in Eq.~\eqref{eq:ZRL} \cite{Antenucci15b,Antenucci16,Antenucci16b}. This is fully equivalent to sample the following probability distribution
\begin{equation} \label{ProbDistr}
P[\bm \hat{\bm{a}}] \propto e^{-\mathcal{H}[\hat{\bm{a}}] }~\delta\left( \mathcal P N - \sum_{k=1}^N |\hat{a}_k|^2 \right),
\end{equation}
where $\bm \hat{\bm{a}}$ denotes rescaled mode amplitude variables $\hat{a}_k \equiv a_k/T^{1/4}$.

So far we have stressed all the formal analogies between the
ML $4$-phasor model 
and other models showing localization:
continuous locally unbounded variables and the presence of a global
constraint. The main difference between
the $4$-phasor partition function (\ref{eq:ZRL}) and the partition function  (\ref{eq:semican-bec}) of the DNLSE  or of the free-bosons, is that in the random laser case
the joint distribution of the variables over which the global constraint
is imposed is not factorized. Therefore,  the analytical
results discussed in~\cite{Gradenigo21a,Gradenigo21b} cannot be straightforwardly extended to this
case. Nevertheless, those systems share a    feature that turns out to be crucial 
also for the present model: the existence of an anomalous  ``pseudo-localized''
phase.  In the non-interacting systems the anomaly is accompanied by a negative temperature, that is,  by a lack of  the canonical-microcanonical ensembles 
equivalence. 
The anomalous phase shows some signatures of incipient localization
but it is  not  a localized phase. As we are going to show, these are the signatures
that we find also in the
ML $4$-phasor model, by means of Monte Carlo numerical simulations above the pumping rate threshold $\mathcal P_c$ (that is proportional to the  intensity threshold $\mathcal \epsilon_c$ at fixed temperature) or, equivalently, below the critical temperature $T_c$, keeping  the spherical constraint fixed ($\epsilon =1$).

\section{Pseudo-localization}

First of all let us provide a qualitative information
about the behaviour of the spectrum when $T$ is lowered, which can be
read off equivalently as an increase of the spherical constraint
value: see Eq.~\eqref{ProbDistr}. In
Fig.~\ref{fig:thermal_spectrum_fpbc} the time-average of the
spectrum at equilibrium is shown for different values of the temperature and a
single instance of the quenched disorder (a single realization of the couplings $J$). It can be very clearly seen that, as the pumping is increased (temperature is decreased) the overall intensity tends to be heterogeneously distributed among the modes. This might hint that a localization phenomenon in intensity occurs but 
it is not enough to establish it.  Whether  the system truly localizes or not can be ascertained only
from the study of the localization order parameter, {the participation ratio}:
\begin{figure}[t!]
\center
\includegraphics[width=.7\textwidth]{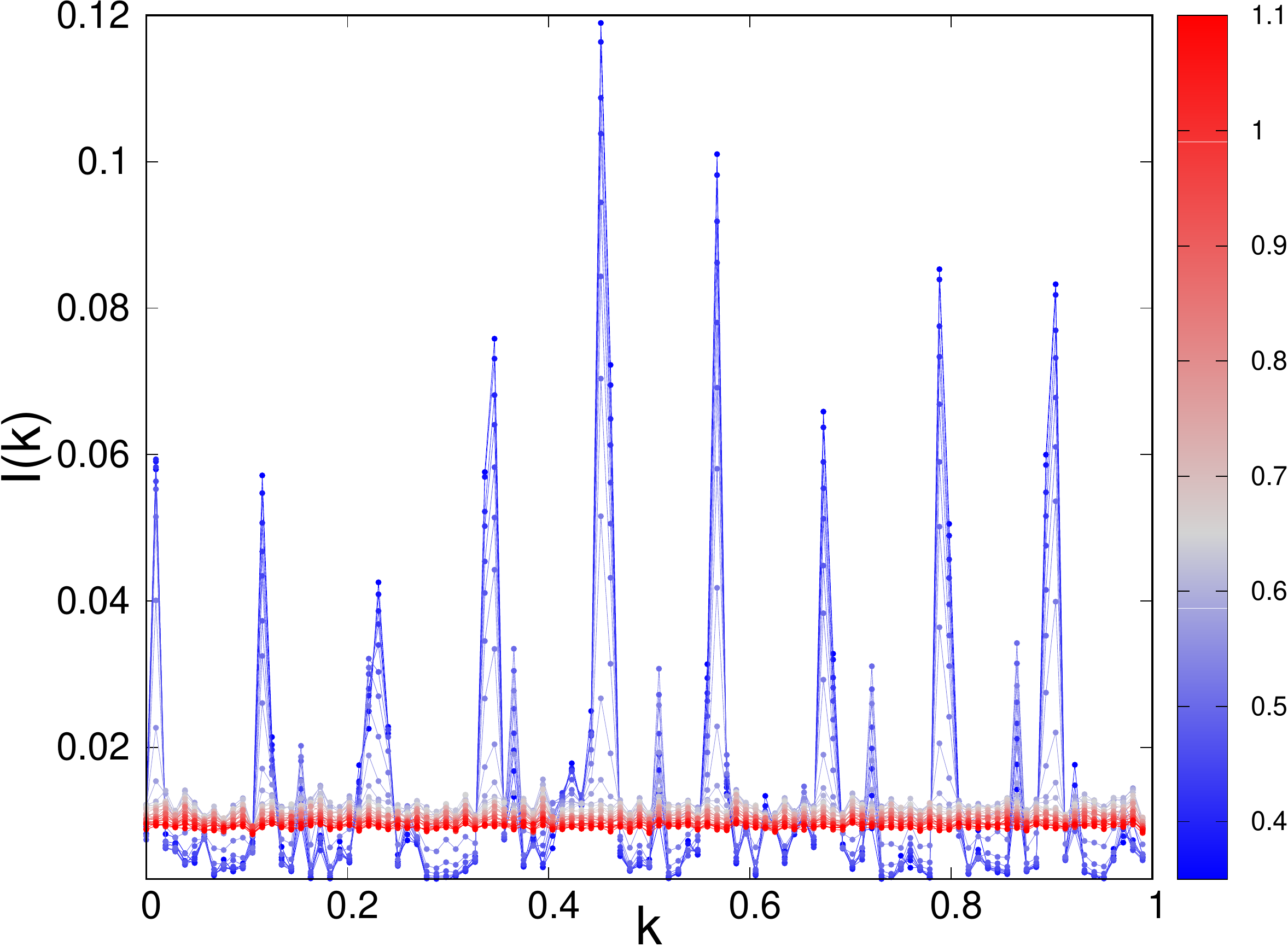}
\caption{Intensity spectra $I(k)=|a_k|^2/T$ for varying optical pumping rate $\mathcal P=1/\sqrt{T}$ ($\epsilon = 1$), see Eq. (\ref{eq:spherical-ok}), as a function
  of the mode index $k$ for a single realization of quenched disorder
  of the ML 4-phasor model with PBC on the frequencies. Numerical
  simulations of size $N=104$. The spectra are normalized and the
  modes $k$ are divided by $N$. Color map: temperature $T$ increases
  (optical power decreases) from blue to red. Notice the absence of
  the spectrum curvature, due to PBC on the FMC, and the presence of
  the isolated peaks below $T_c\simeq 0.61$.}
\label{fig:thermal_spectrum_fpbc}
\end{figure}
\begin{equation} 
  Y_2 = \left\langle \frac{\sum_{k=1}^N {I}^2_k
  }{\left(\sum_{k=1}^N {I}_k\right)^2} \right\rangle =\frac{1}{N^2} \left\langle
  \sum_{k=1}^N {I}^2_k
  \right\rangle, 
    \label{Y2amp}
\end{equation}
where ${I}_k = |a_k|^2$ and we have used that  
\begin{equation}
\sum_{k=1}^N {I}_k = N
\end{equation}
 because of equation (\ref{spherConstr}), with $\epsilon=1$.  
The dependence on the number of degrees of freedom of $Y_2$ can be
easily rationalized in two extreme situations: equipartition and
localization of the mode intensity. Let us consider localization first: in this case a
finite fraction of the whole intensity is taken by  a
finite number of modes that does not increase with  $N$, see Appendix \ref{app:loc_vs_equ}. That is, in the localized phase, a few modes $k$ have intensity
\begin{equation}
{I}_k \propto N,
\end{equation}
whereas all the others $ I_{\neq k}= 0$. 
Then, we have 
$$\sum_{k=1}^N I^2_k\simeq \sum_{k=1}^{\# {\rm loc\  modes}} \hspace{-.3cm} I^2_k\propto N^2.$$
This implies that in a localized phase the participation ratio $Y_2$  in
the limit $N\rightarrow \infty$ is a constant that does not depend on
$N$:
\begin{equation}
\text{localization}~\Longleftrightarrow~  \lim_{N\to\infty }Y_2 =\mbox{const}.
\label{LOCALIZATION_Y}
\end{equation}
On the contrary, in the equipartite phase of nearly homogeneous spectral intensities, any of the $N$ modes has intensity
${I}_k=\mathcal{O}(1)$, so that in the thermodynamic limit  

\begin{equation}
\text{equipartition}~\Longleftrightarrow~ Y_2 \sim \frac{1}{N}.
\end{equation}

\begin{figure}[t!]
\center
\includegraphics[width=.8\textwidth]{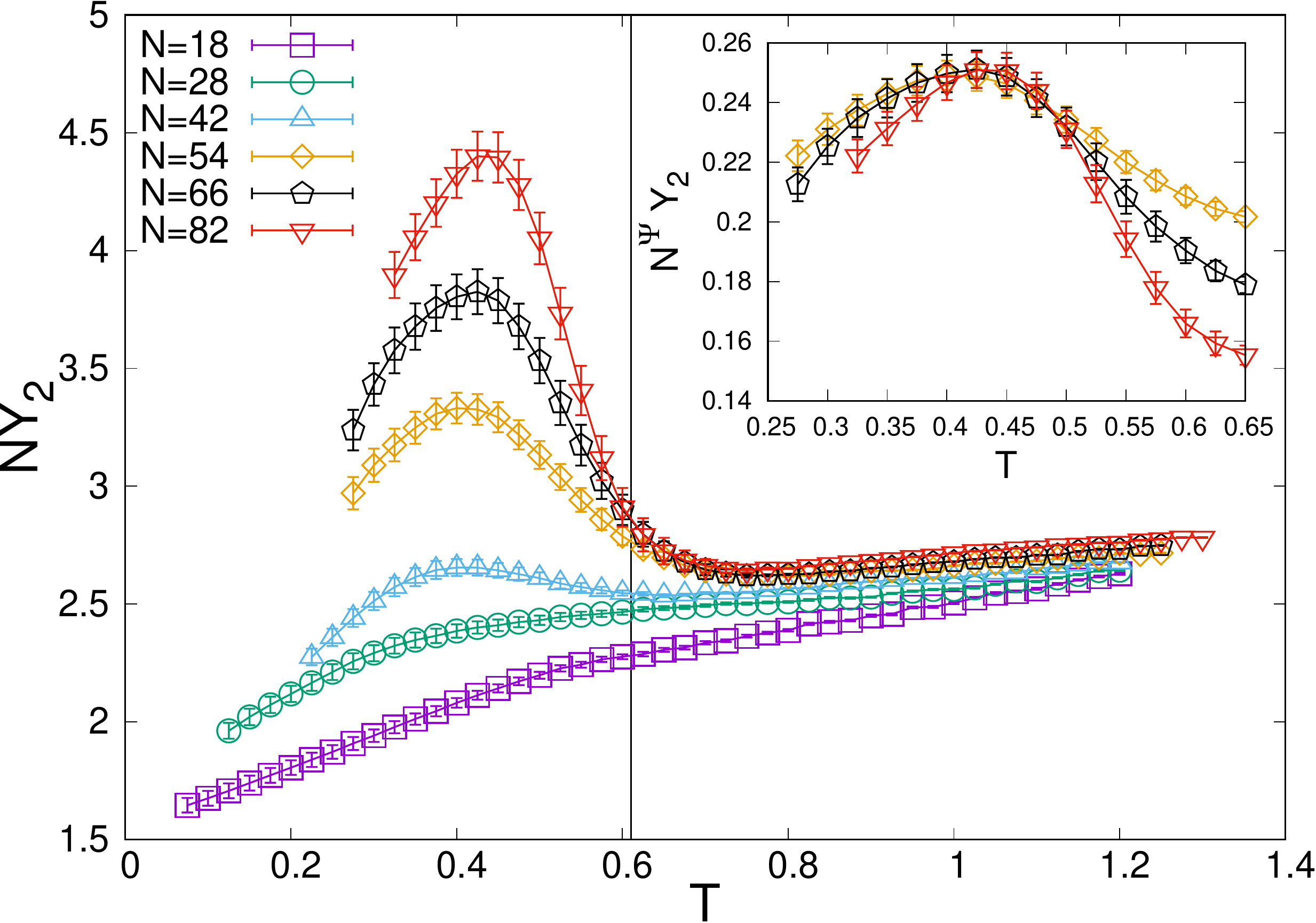}
\caption{Main: Participation ratio $NY_2$ of the mode intensities versus $T$ at different systems sizes $N$. The vertical line is the asymptotic value of the critical temperature for the system $T_c\simeq 0.61$ (from \cite{Niedda22}). Inset: scaling of the participation ratio near the peak for the three largest sizes, as $N^\Psi
  Y_2$ versus $T$, with $\Psi=0.35>0$. The peak height scaling is less than $N$, thus Eq. (\ref{Y2amp}) tends to  zero as $N\to\infty$.}
\label{fig:Y2-ampiezze}
\end{figure}

Now we are ready to display,  in
Fig.~\ref{fig:Y2-ampiezze}, the first important quantitative information
obtained from the study of the equilibrium distribution of the
intensity among modes. In the
figure we have plotted for convenience the average over quenched disorder of $ N Y_2(T)$, which we
expect of $ \mathcal{O}(1)$ in the equipartite phase and of
$\mathcal{O}(N)$ in a possible localized phase.
We observe that in the high temperature phase $NY_2\sim \mbox{const}$ and, therefore,  the system is in the 
equipartition regime. 
Under the critical point, indicated by the vertical line at $T_c=0.61$, which is the glass transition temperature {\color{black}{for this specific model extrapolated inthe thermodynamic limit}} (see \cite{Niedda22}), we find, instead, {\color{black}{a departure from equipartition. Such behaviour is evident in the whole low T region probed by our numerical simulations. Yet, $NY_2(T)$ does not display an extensive scaling with $N$, as one would expect for the localized regime according to Eqs. (\ref{Y2amp},\ref{LOCALIZATION_Y}). In the
main panel of Fig.~\ref{fig:Y2-ampiezze}, by collapsing the curves around $T=0.42$, where the peaks of $NY_2(T)$ occur, we observe that   $NY_2(T)$ grows with $N$  less than linearly.
In this specific case we find  $Y_2(T)\sim N^{-\Psi}$
with $\Psi \simeq 0.35$.  This is the slowest scaling obtained from our data.
As lower $T$ regions are considered for the scaling with $N$ the $\Psi$ exponent slightly increases  in the range of simulated temperatures and sizes (it is $\Psi\simeq 0.5$ at the lowest simulated temperatures).
We, thus, find no evidence for the low $T$ regime to show intensity condensation, though  it is not equipartite either. }}

The difference from the localization scaling in $N$, cf. Eq. (\ref{LOCALIZATION_Y}), i.e. $\Psi=0$, cannot be accounted for as a finite size effect, as it is usually hypothesized in the estimate of critical exponents of a second order phase
transition. The latter effects are, indeed,  due to  the cutting
of long-wavelength fluctuations in a finite size
 lattice. Localization is, instead, controlled by a first-order
mechanism where a finite fraction of the whole localizing
quantity concentrates on a few variables in such a way that
$Y_2$  is strictly independent from $N$.
In the glassy phase we, thus, have a
phase that might show some signature of incipient localization but it
is certainly not localized in intensity. Intensity equipartition is broken but no finite group of modes takes all the intensity of the system.
A scaling argument for the onset of such phenomenon is reported in Sec. \ref{ss:connectivity} and Appendix \ref{app:loc_vs_equ}.

The existence of a similar pseudo-localized phase
 has  been predicted for the DNLSE~\cite{Gradenigo21a,Gradenigo21b}. The peculiarity of the localization
phenomenon in that case is that it takes place in
two steps. By increasing the energy, i.e.~the  quantity  controlling localization, one first encounters a second order
transition at a value of the energy --  $E_{\rm th}$ --  where the
equivalence of ensembles breaks down and temperature becomes negative.
Then, at a larger value of energy, $E_c > E_{th}$, there is a
first-order transition to a localized phase. As we mentioned already
in the random laser case the quantity  controlling localization is the overall light mode
intensity, and not the energy, so that we cannot identify the transition to
the pseudo-localized phase with a negative temperature and the breaking of
equivalence between canonical and microcanonical ensembles.

What
can  signal the onset of a non-trivial pseudo-localized phase in the ML $4$-phasor model?
A straightforward answer
is to look for an indicator of equipartition, {\color{black}{like the spectral entropy or the related  effective fraction}} of degrees of freedom \cite{Livi85,Cretegny98}.

The spectral entropy is defined as 
\begin{equation}
S_{\text{sp}} = - \sum_{k=1}^N \hat{\mathcal{I}}_k \ln(\hat{\mathcal{I}}_k),
\end{equation}
where $\hat{\mathcal{I}}_k$ is the thermal averaged intensity of the
mode $k$ normalized to the total intensity of the spectrum
\begin{equation}
\hat{\mathcal{I}}_k = \frac{\langle {I}_k \rangle}{\sum_{k=1}^N \langle {I}_k \rangle} = \frac{\langle |a_k|^2 \rangle}{N\epsilon}.
\end{equation}
\begin{figure}[t!]
\center
\includegraphics[width=.8\textwidth]{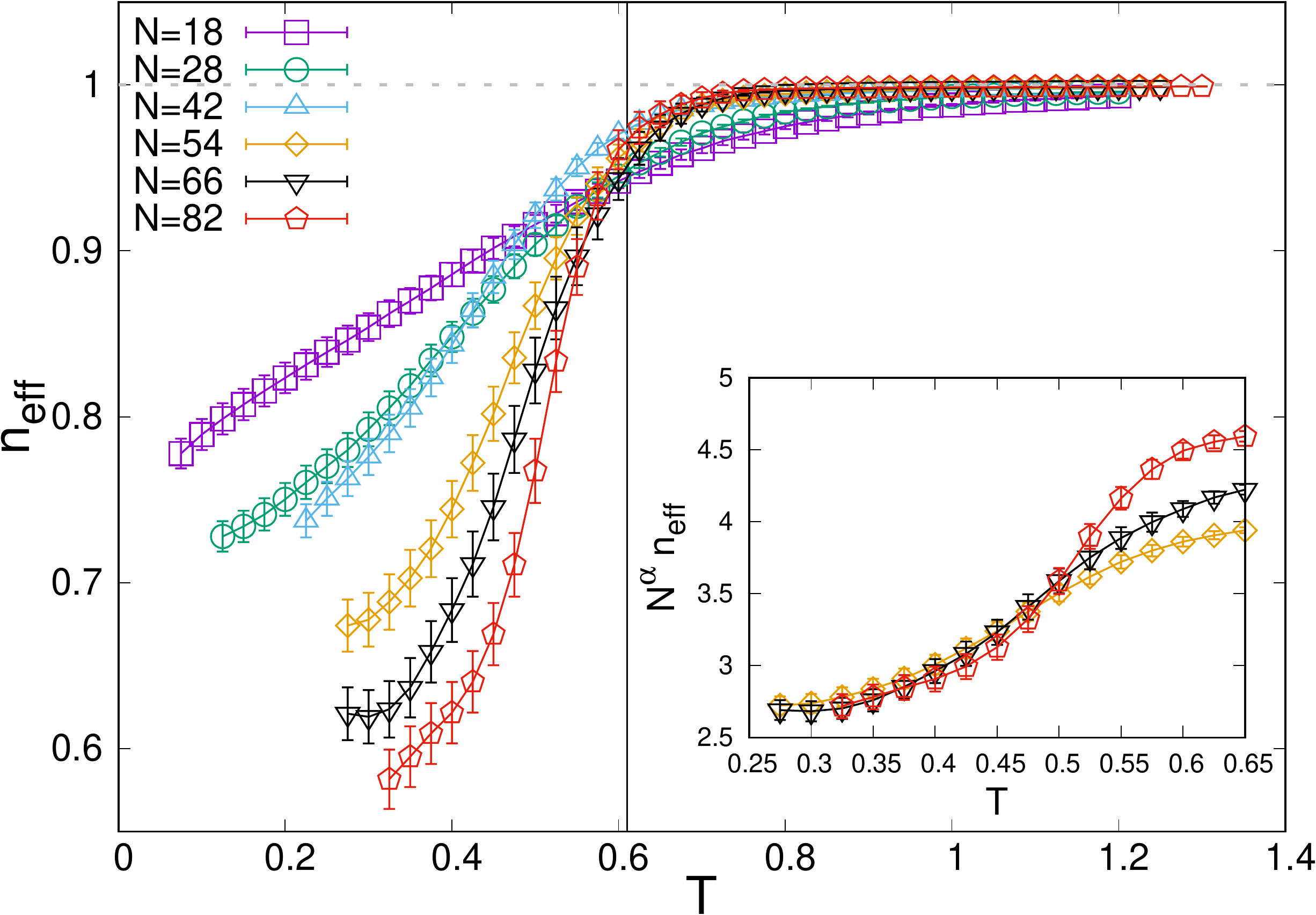}
\caption{Main: Effective number of degrees of freedom $n_{\text{eff}}$ versus $T$ for the mode
  intensities $|a_k|^2$ for different systems sizes. The vertical line denotes the critical temperature for the system, $T_c\simeq 0.61$. Inset: $N^\alpha n_{\rm eff}$ versus $T$, $\alpha=0.33$.}
\label{fig:Neff-ampiezze}
\end{figure}
The effective {\color{black}{fraction}} of degrees
of freedom, which is a function of the spectral entropy and  {allows to better visualize the information encoded in the spectral entropy}, is defined as
$$
n_{\text{eff}} = \frac{e^{S_{\text{sp}}}}{N} .$$
The behaviour of $n_{\text{eff}}$, averaged over quenched disorder, is reported in
Fig.~\ref{fig:Neff-ampiezze} and shows the signature of a 
phase transition, where equipartition breaks down, at {\color{black}{a temperature compatible with the critical transition
    temperature of the glassy random laser with frequency periodic boundary condition.  Its behavior turns out to be qualitatively similar to the one of the effective fraction of degrees of freedom for the model with free boundary conditions on the frequencies analyzed in Ref.~\cite{Gradenigo20}. 
    The same analysis with further indicators shows that}} this equipartition-breaking transition is of first order in the parameter  $n_{\text{eff}}$.

In Fig.~\ref{fig:Neff-ampiezze} we notice that in the high temperature phase all the curves
 approach one, while they decrease to a size-dependent
quantity for low temperature: the larger the size, the steeper
${n}_{\text{eff}}$ decreases. In the inset, we plot the rescaled
effective fraction of degrees of freedom for the three  largest sizes. In the low temperature phase they collapse on each other with an exponent
$\alpha\simeq 1/3$. Thus, in the thermodynamic limit, $n_{\rm eff}$ tends
to $0$ below the RSB critical point, marking the breaking of
equipartition.

The transition taking place at $T_c$, besides being a glass transition, can, thus, be 
characterized as the transition to a phase with a thermodynamic
anomaly, the breakdown of equipartition, in an analogous way
 for which we have a breakdown of ensemble equivalence in a non-interacting system such as 
the DNLSE. 

\begin{figure}[t!]
\center
\includegraphics[width=0.8\textwidth]{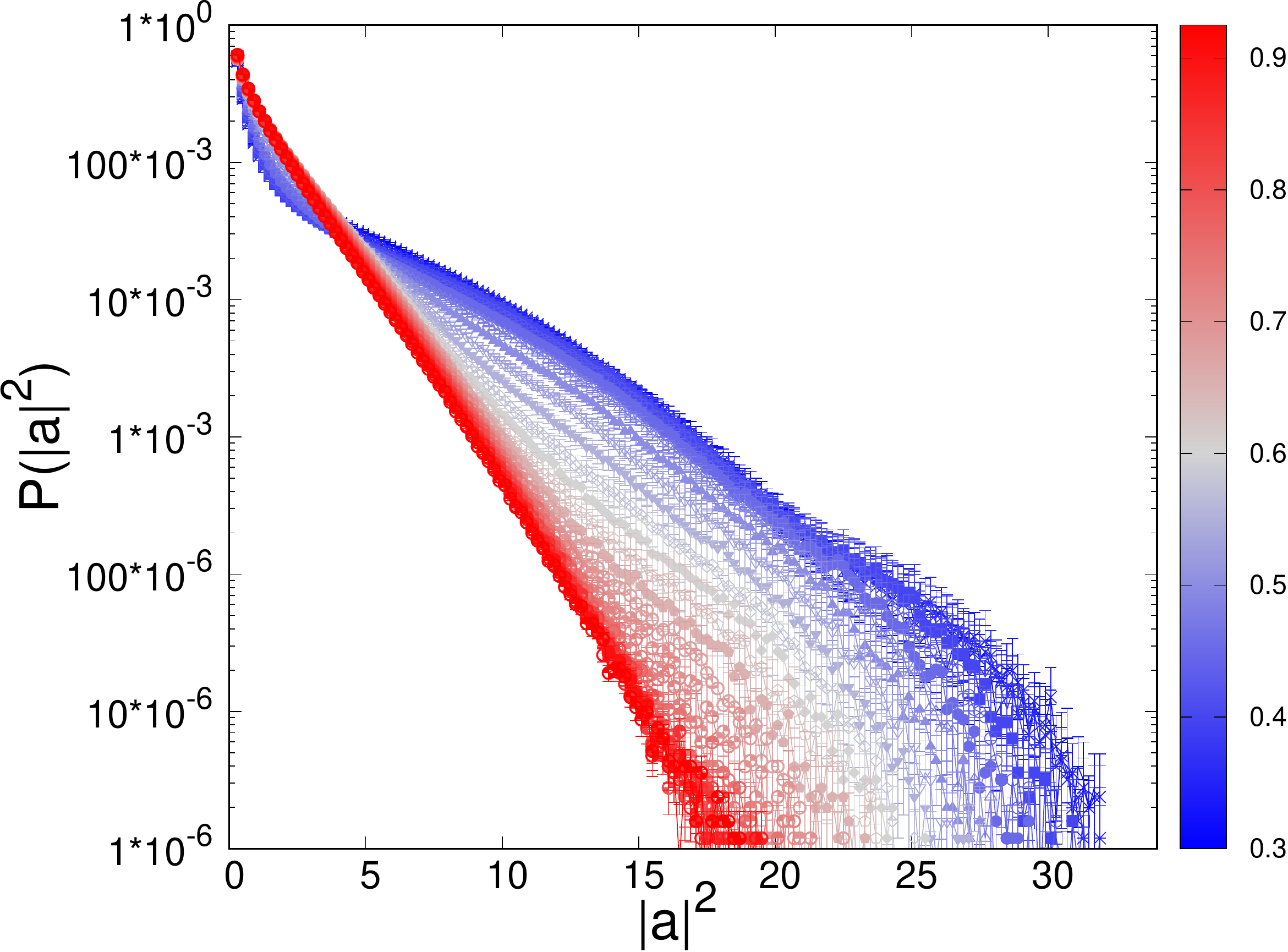}
\caption{Marginal distribution of spectral intensity averaged over quenched disorder for the system size $N=82$ at all the simulated temperatures. Color map: as in Fig.~\ref{fig:thermal_spectrum_fpbc}. At high temperature the distribution is exponentially decaying as revealed by the linear trend in semilogarithmic scale (red curves). By lowering the temperature deviation from monotonicity can be observed (blue curves) in relation to the onset of the pseudo-localized phase.}
\label{fig:marginal-1}
\end{figure}

\begin{figure}[t!]
\center
\includegraphics[width=0.8\columnwidth]{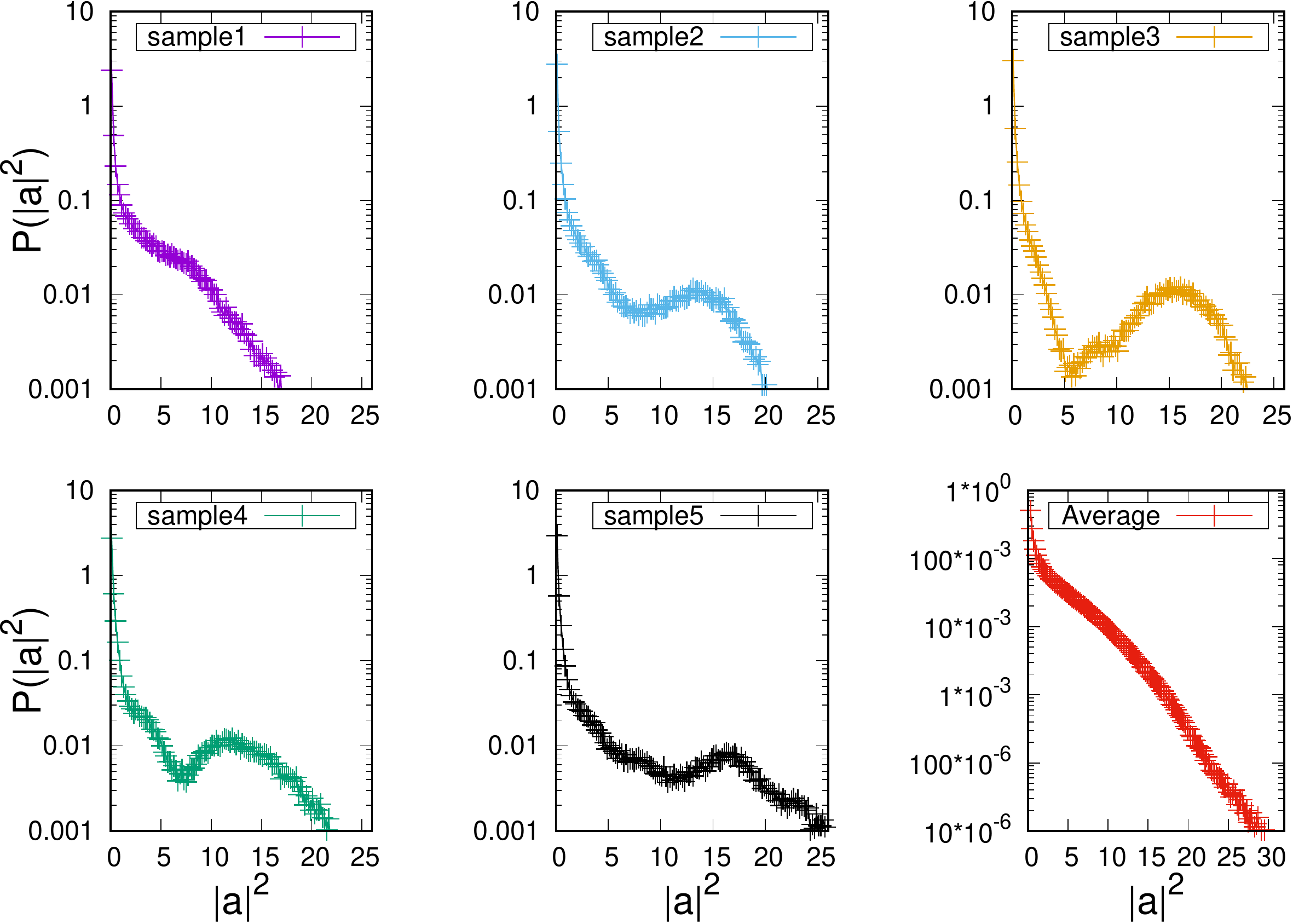}
\caption{Marginal intensity distributions for single instances of the quenched disorder. Data are referred to the simulated size $N=82$ at the temperature $T=0.3$.
We notice that some of the samples (such as sample 2,3,4 and 5) exhibit a  peak for large mode intensity of the distribution, corresponding to accumulation of the intensity on a few modes. 
The position of the peak is sample dependent. On the other hand, other samples behave as sample $1$, exhibiting only thick tails in the marginal intensity distribution.  The result of averaging over all the disordered samples is reported in the sixth panel, which corresponds to the lowest temperature intensity distribution in Fig.~\ref{fig:marginal-1}.}
\label{fig:marginal-2}
\end{figure}

{\color{black}{
The analogy with the  DNLSE holds also for what concerns the spectral intensity distributions, cf. Figs. \ref{fig:marginal-1}, \ref{fig:marginal-2}.
We can observe in Fig.~\ref{fig:marginal-1} that, as temperature is lowered, the occurrence of the anomalous phase with lack of
equipartition is
signaled by the 
onset of fat tails of the quenched average distribution $P(|a|^2)$. }}
{\color{black}{
By looking at the distributions for single instances of the random couplings  $J$, see Fig.~\ref{fig:marginal-2}, one can understand that the tails in the average distribution come about because of the  occurrence of a secondary peak at large intensity, which, 
in a hypothetical strictly condensate phase, would become the typical
``condensate bump''. 
This peak  also appears in the marginal distribution of energy in the DNLSE (see Fig.~5 in~\cite{Gradenigo21a}). In that case, though, the  non-monotonic behavior of the marginal intensity distribution eventually signals the approach to a truly localized phase.}}


\section{Conclusions}

In this work we have presented a detailed analysis of which signatures of intensity localization in light modes spectrum are emerging in the glassy phase of light when the interaction between modes is represented by the so-called ``mode-locked" $p$-spin model {\color{black}{with periodic boundary condition on the mode frequencies. }} This model offers a unique benchmark to study the coexistence between localization and phase-space fragmentation, thanks to its peculiarity with respect to the most commonly studied models of disordered systems: 
{\color{black}{it has continuous, locally unbound, variables with non-linear diluted interactions. These are diluted enough  to allow for relevant intensity heterogeneities at low $T$, cf. Fig~\ref{fig:thermal_spectrum_fpbc}.}}

From a careful finite-size study of the participation ratio of the mode intensities
we have been able to assess that, although some signatures of \emph{incipient} localization can be found, the glassy phase of light is not, strictly speaking, localized in intensity.
This means that the results of our analysis are not compatible with single light modes carrying an extensive amount of intensity, i.e.~there is no $k$ such that $|a_k|^2 \sim N$. 
We have found an anomaly in the finite-size study of the participation ratio, whose size dependence is compatible with the presence of high power modes, carrying an intensity $|a_k|^2 \sim N^{1-\Psi/2}$, with $\Psi > 0$.
We stress that in presence of power condensation those modes would have intensity
$|a_k|^2 \sim N$, cf. Eq. (\ref{LOCALIZATION_Y}), i.e., $\Psi=0$.
 We have termed this phase ``pseudo-localized" phase. 
 
 In recent literature~\cite{Gradenigo21a,Gradenigo21b},  a similar phase has been discovered and characterized from the theoretical point of view in a simpler non-interacting model,
 the discrete non-linear Schr\"odinger equation (DNLSE).  
 In both models the pseudo-localized phase is characterized by a non-monotonic shape of the marginal probability distribution of the  quantity under probe, i.e., 
 the site energy for the DNLSE and the overall intensity of light modes for the mode-locked $p$-phasor model.
 Due to the different nature of the global constraint controlling  the localization and  to the presence of mode interactions in the random laser model,{\color{black}{ though, the mechanisms leading to the pseudo-localized phase differ.}} The DNLSE is characterized by negative temperature, whereas the pseudo-localized phase of the mode-locked random laser model is characterized by the breaking of equipartition, which can be observed very clearly in emission spectra upon increasing the pumping rate both in numerical simulations and experiments on random lasers \cite{Mujumdar:07}.  
The parameter driving the localization transition in the DNLSE is the energy, while in the random laser power it appears to be the dilution of the interaction network. 

In conclusion, we have presented a  comprehensive study of light intensity localization (on Fourier modes), also known as \emph{power condensation} \cite{Antenucci15c,Antenucci15d}, in a disordered model for glassy light{\color{black}{, the mode-locked random laser with periodic boundary condition on the frequencies \cite{Niedda22}.}} There are still many open questions and problems to be addressed: in the first place, to achieve an analytic control of the localization phenomenon in non-trivial disordered interacting models like the mode-locked $p$-phasor model; second, by means of a consistent theoretical modelling of the phenomenon, to gather a precise quantitative control of the dilution needed to allow the appearance of a truly localized phase in the model.

\section{Acknowledgements}
We thank Daniele Ancora, Maria Chiara Angelini, Roberto Livi, Giorgio Parisi and
  Federico Ricci Tersenghi  for useful discussions.
We acknowledge the support of LazioInnova - Regione Lazio under the program Gruppi di ricerca 2020 - POR FESR Lazio 2014-2020, Project NanoProbe (Application code A0375-2020-36761).

\appendix
\section{Intensity equipartition vs localization and the interaction network connectivity}
\label{app:loc_vs_equ}
Let us consider a generic $p$-spin interacting model with $N$ continuous spins $\sigma$ and a spherical constraint 
\begin{equation}
    \sum_{i=1}^N \sigma_i^2 =N,
    \label{eq:SpherCon}
\end{equation} whose Hamiltonian is
\begin{equation}
    \mathcal H[\sigma] = -\sum_{k_1\ldots k_p}^{\# N^A} \sigma_{k_1}\ldots \sigma_{k_p} \ J_{k_1\ldots k_p}, 
    \label{eq:Hp}
\end{equation}
where $N^A$ on top of the sum, with $A\in [1,p]$,
 denotes the scaling with the size of the number of $p$-uples contributing to the energy and the spin indices  $k_{i}$ run from $1$ to $N$. 
 If $A=p$ we have a fully connected interaction graph, i.e.,~each spin contributes in $\mathcal O(N^{p-1})$ $p$-uples. 
 On the other extreme, if $A=1$ the graph is {\em sparse}, i.e.~each spin only interacts in a finite  number of $p$-uples, not growing with the size of the system. 
 All dilutions in between will be considered hereafter. 
 
 For simplicity, we will take the variables as real valued here, yet  keeping the word {\em intensity} for the spin magnitude $|\sigma|$.

 We notice that 
the glassy random laser is a model belonging to this family, with $p=4$ but with complex spins. Though complex variables yield some new physical features we stress that  these are not relevant for what concerns the influence of connectivity on the possible onset of intensity localization. 

First we consider the ordered case $J_{k_1\ldots k_p}=J$. The typical ground state spin configurations in the canonical ensemble, where equipartition is expected to hold (in the average, not strictly), are those minimizing Eq.~(\ref{eq:Hp}). The energy is extensive
$$E=\mathcal H[\sigma_{\rm gs}]=\mathcal O(N)$$
provided that the coupling constant scales as \begin{equation}
    J\propto \frac{1}{N^{A-1}}.
    \label{eq:JNscale}
\end{equation}

If intensity localization occurs, that is, if only a few modes take the overall intensity, equal to $N$ according to  Eq.~(\ref{eq:SpherCon}), whereas all the other are zero,  what will be the energy contribution of a localized spin configuration?
First of all let us notice that in order to have a non-zero contribution the intensity of at least  $p$ coupled spins must localize. If we represent by $\square$ such a localizing $p$-uple the  intensity localized configuration is
\begin{equation}
    \{\sigma_{\rm loc}\}: \qquad \sigma \in \square \propto \sqrt{N} \quad , \quad  \sigma \cancel{\in} \ \square =0.
    \label{eq:localization}
    \end{equation}
According to Eqs. (\ref{eq:Hp}), (\ref{eq:JNscale}) the energy of such a configuration of spins scales with $N$ like 
$$E_{\rm loc}=\mathcal H[\sigma_{\rm loc}] = 
\mathcal O\left(\frac{N^{p/2}}{N^{A-1}}\right).$$

To figure out whether intensity condensation might occur and dominate, one eventually has to compare the scaling behaviors of the energies of an equipartite and a localized configuration:
$$\mathcal O(N) \quad \mbox{vs}\quad  \mathcal O(N^{\frac{p}{2}+1-A}).$$

The following cases occur depending on the interaction connectivity scaling $N^A$:
\begin{enumerate}
    \item $A>p/2$. Any possible intensity localized configuration of spins would yield subextensive contributions to the energy. The equipartite regime is, therefore, dominant. The case $A=p$ is the fully connected interaction graph.
    
    \item $A=p/2$. Both kinds of spin configurations yield an $\mathcal O(N)$ contribution to the energy. In this case a pseudo-localized phase might occur. 
    
    \item $A<p/2$. Intensity localization provides the most prominent contribution to the energy, that is, $\mathcal O(N^{>1})$. The case $A=1$ is the sparse case. 
\end{enumerate}

If the interaction couplings $J_{k_1\ldots k_p}$ are quenched disordered, independently distributed and with zero mean ${\overline{J_{k_1\ldots k_p}}}=0$, the typical ground state of the Hamiltonian (\ref{eq:Hp}) is extensive -- $E=\mathcal H [\sigma_{\rm gs}]=\mathcal O(N)$ --  if the variance of the distribution of the couplings scales like
\begin{equation}
    \label{eq:JNdisscale}
{\overline{J^2_{k_1\ldots k_p}}}\propto \frac{1}{N^{A-1}}.
\end{equation}
If the total intensity of the system is localized in a single interacting $p$-uple, as in (\ref{eq:localization}), Eqs. (\ref{eq:Hp}), (\ref{eq:JNdisscale}) imply that the energy scales with the size like
$$E_{\rm loc}=\mathcal H[\sigma_{\rm loc}] = 
\mathcal O\left(\frac{N^{p/2}}{N^{(A-1)/2}}\right).$$
Comparing  the equipartite contribution $E=\mathcal O(N)$ and the localized contribution $E_{\rm loc}$
we find the following three regimes as the exponent $A$ varies:

\begin{enumerate}
    \item $A=p$. Localized energy contibutions are subextensive. The equipartite regime is dominant. This  is the fully connected interaction graph case.
    
    \item $A=p-1$. Both kinds of spin configurations yield an $\mathcal O(N)$ contribution to the energy. This is the case, e.g., of the mode-locked glassy random laser, studied in the present work, where $p=4$, $A=3$.
    In this case one might conjecture the occurrence of a pseudo-localized phase. 
    
    \item $A<p-1$. Intensity localization provides superextensive contributions to the energy.  
    
\end{enumerate}
\section{Monte Carlo simulation algorithm on GPU's}

The numerical simulations of the ML 4-phasor model have been performed
by means of a Parallel Tempering Monte Carlo algorithm~\cite{Hukushima96} parallelized on GPU?s. All the details about the algorithm can be found in~\cite{Gradenigo20,Niedda22}. Here, we just sketch its most
important features and report the details of the simulations, for the
reader's convenience.

\begin{table}[t!]
\centering
\begin{tabular}{lrrrrrr}
\hline \hline
$N$ & $N_4$ & $T_{\text{min}}$ & $T_{\text{max}}$ & $N_{\text{PT}}$  & $N_{\text{MCS}}$ & $N_{\text{s}}$ \\
\hline 
18 & $2^{9}$  & 0.05 & 1.2 & 46 & $2^{19}$ & 200 \\
28 & $2^{11}$  & 0.1 & 1.2 & 44 & $2^{19}$ & 200 \\
42 & $2^{13}$  & 0.2 & 1.2 & 40 & $2^{20}$ & 150 \\
54 & $2^{14}$  & 0.25 & 1.25 & 40 & $2^{20}$ & 100 \\
66 & $2^{15}$  & 0.25 & 1.25 & 40 & $2^{20}$ & 100 \\
82 & $2^{16}$  & 0.3 & 1.3 & 40 & $2^{20}$ & 100 \\
104 & $2^{17}$ & 0.35 & 1.3 & 38 & $2^{21}$ & 80 \\
\hline \hline
\end{tabular}
\caption{Details for the numerical simulations of the ML 4-phasor model with PBC on the frequencies.}
\label{tab1}
\end{table}

Given a system size $N$, for each sample of disorder, the Mode-Locked graph of the
interactions is generated, by applying the Frequency Matching Condition with frequency Periodic Boundary Condition to a generated
fully-connected graph with $N(N-1)(N-2)(N-3)/24$ quadruplets. Each one of these, if  satisfying the FMC,
is progressively added to the ML graph up to a preassigned number $N_4$: the power of $2$ nearer to the 
 total number of quadruplets satisfying the FMC at the given simulated size. 
 
The numerical values of the interacting quadruplets
$\{J_{\bm{k}}\}$ are independently extracted from the Gaussian
distribution Eq.~\eqref{Gaussian}. The Metropolis dynamics of
$\text{NPT}$ replicas of the system at different temperatures is run
in parallel on GPU?s with the Parallel Tempering algorithm. The temperatures are chosen in order to have
almost one third of the replicas below the critical temperature and
two thirds above and have an exchange rate between thermal baths at nearby temperatures staying constant for all temperature couples.

A subtle point of our algorithm is that each update of the dynamics
must be compatible with the spherical constraint $\sum_k |a_k|^2
=\epsilon N$. Given the non-locality of this constraint on the
configurations, variables updates must be done sequentially. Given two
randomly chosen variables $a_i = A_i e^{i\phi_i}$ and $a_j=A_je^{i\phi_j}$, 
the local attempt is based on the extraction of three random numbers 
$x,y \in [0,2\pi]$ and $z \in [0, \pi/2]$: the first two correspond to the 
new (attempted) phases $x=\phi_i'$ and $y=\phi_j'$, 
the third one mixes the intensities of the modes selected 
by preserving the spherical constraint, since
$A_j' = \sqrt{A_i^2 + A_j^2} \cos z$ and $A_j' = \sqrt{A_i^2 + A_j^2} \sin z$.
The computation of the energy shift between the original configuration 
and the attempted one ($a_\ell \to a'_\ell$) requires a number of operations  scaling with
 the number of quadruplets involved in the variation of $a_\ell$
\begin{eqnarray}
\nonumber
    \Delta E_\ell &=& \left(a_\ell-a_\ell'\right)
    \hspace*{-.5cm}\sum_{k_1,k_2,k_3}^ {{\rm FMC}(k_1,k_2,k_3,\ell)}\hspace*{-.5cm} J_{k_1k_2k_3\ell}\  \bar a_{k_1} a_{k_2} \bar a_{k_3} + \mbox{c.c.} 
    \\
    \nonumber
    &=& \mathcal{O}(N^2).
\end{eqnarray}
This computation has been parallelized on GPU?s, to speed up the algorithmic time and perform the above average in $\mathcal O(\ln N)$ steps.
Moreover, also the Parallel Tempering dynamics at different temperatures has been parallelized on GPU's. 
The two kinds of parallelization considered together 
reduce the execution time of the entire simulation by a factor of $8$.

Exchanges among configurations of replicas at adjacent temperatures
are proposed after a fixed number of steps of the local Metropolis
algorithm and accepted with a probability that depends on the
Boltzmann weights of the replicas involved in the exchange.
This helps to speed up the relaxation dynamics of the low
temperature replicas of the system, which move in a complex rugged
free-energy landscape. We rigorously ascertained the thermalization of
each sample by looking at the relaxation of energies on logarithmic
time windows and by checking the symmetry of the Parisi overlap
distribution function, the order parameter of the glass
transition~\cite{Mezard87}.

Once thermalization is tested and an equilibrium time 
$\tau_{eq}$ is estimated, we sampled $\mathcal N = (N_{\text{MC}} -
\tau_{eq})/\tau_{\text{corr}}$ equilibrium configurations, where
$N_{\text{MC}}$ is the total number of Monte Carlo steps chosen large
enough to have a good statistics and $\tau_{\rm corr}$ is the auto-correlation
time of mode amplitudes, that  we consider in order to not underestimate statistical errors. Canonical
ensamble averages over observables are computed through the time
average over uncorrelated equilibrium configurations $\bm{a}_t$
\begin{equation}
\langle O[\bm{a}] \rangle = \frac{1}{ \mathcal N}\sum_{t=\tau_{\rm eq}/\tau_{\rm corr}}^{N_{\rm MCS}/\tau_{\rm corr}} O[\bm{a}_t].
\end{equation}

All details of the simulations are reported in Table~\ref{tab1}. 

\providecommand{\newblock}{}

\end{document}